\documentclass{egpubl}
\usepackage{eg2026}

\usepackage[T1]{fontenc}
\usepackage{dfadobe}
\usepackage{algorithm}
\usepackage{algpseudocode}
\usepackage{amsmath}
\usepackage{amsfonts}
\usepackage{bm}
\usepackage{bbm}

\usepackage{cite}  
\BibtexOrBiblatex
\electronicVersion
\PrintedOrElectronic
\ifpdf \usepackage[pdftex]{graphicx} \pdfcompresslevel=9
\else \usepackage[dvips]{graphicx} \fi

\usepackage{egweblnk} 
\usepackage{lineno}
\usepackage[dvipsnames]{xcolor}

\usepackage{pgfplots}
\pgfplotsset{compat=1.18} 

\usepackage{tikz}
\usetikzlibrary{arrows.meta,calc,decorations.markings}

\newcommand{\mat}[1]{\bm{#1}}

\newcommand{\Rn}[1]{\mathbb{R}^{#1}}

\newcommand{\Nn}[1]{\mathbb{N}^{#1}}

\DeclareMathOperator*{\argmin}{argmin}

\newcommand{\reffig}[1]{Fig.~\ref{fig:#1}}
\newcommand{\refeq}[1]{Eq.~(\ref{eq:#1})}

\newcommand{\refalg}[1]{Algorithm~\ref{alg:#1}}

\title[]%
      {MorphModes: Non-rigid Registration via Adaptive Skinning Eigenmodes }

\author[]
{\parbox{\textwidth}{\centering G. Browne*$^{1}$ and M. Liu*$^{1}$
        and E. Grinspun$^{1}$ and O. Benchekroun$^1$
         \\ $^1$University of Toronto}}
 
\begin{document}

\teaser{
 \includegraphics[width=0.99\linewidth]{Lion_Lion_Teaser.pdf}
 \centering
  \caption{We introduce a Non-Rigid Registration algorithm that minimizes the SDF error between a source and a target shape.  The optimization is carried out using a novel adaptive Skinning Eigenmode subspace, leading to smooth, stable morphing of one lion pose to another.}
\label{fig:teaser}
}

\maketitle


\begin{abstract}
   Non-rigid registration is a crucial task with applications in medical imaging, industrial robotics, computer vision, and entertainment.
   Standard approaches accomplish this task using variations on the Non-Rigid Iterative Closest Point (NRICP) algorithms, which are prone to local minima and sensitive to initial conditions.
   We instead formulate the non-rigid registration problem as a Signed Distance Function (SDF) matching optimization problem, which provides richer shape information compared to traditional ICP methods. 
   To avoid degenerate solutions, we propose to use a smooth Skinning Eigenmode subspace to parameterize the optimization problem.
   Finally, we propose an adaptive subspace optimization scheme to allow the resolution of localized deformations within the optimization.
   The result is a non-rigid registration algorithm that is more robust than NRICP, without the parameter sensitivity present in other SDF-matching approaches.


\end{abstract}  
\begin{small}
*Both authors contributed equally to this research. \\ \\
\end{small}
\hrule

\section{Introduction}

How does one shape transform into another? 
Registration is the process of finding an answer to this question. 
To an artist, this might seem like a simple task.
For example, mapping between a human body and a quadrupedal creature has an obvious solution: the head of the human transforms into the head of the creature, the legs to the back legs and the arms to the front legs.
However, coming up with an algorithmic solution to this problem is much more challenging. 

The \textit{de-facto} approach to perform registration is via the Iterative Closest Point (ICP) algorithm \cite{besl1992ogipc}.
Unfortunately, despite its widespread use, ICP suffers from well-known limitations.
Most notoriously, ICP, especially its non-rigid variant, is prone to severe local minima and null space problems that arise from the closest point projection that occurs at every one of its iterations; so long as a point on the source shape lies on the target shape, its contribution to the ICP energy will be minimized.
\begin{figure}[t]
  \centering
  \includegraphics[width=\linewidth]{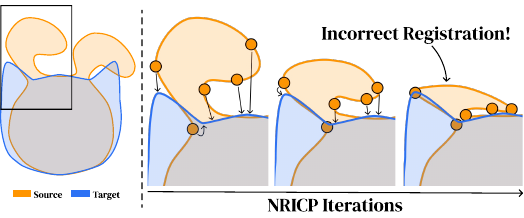}
  \caption{\label{fig:nricp_didactic}
           Didactic illustration of the shrinkage occurring throughout non-rigid ICP iterations leading to incorrect registration.}
\end{figure}
\begin{figure}[b]
  \centering
  \includegraphics[width=\linewidth]{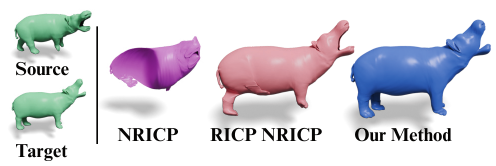}
  \caption{\label{fig:hippo}
           An example showing limb shrinkage occurring with ICP-based registration. Directly using NRICP leads to extreme shape collapse. Aligning the two meshes first using RICP, followed by NRICP leads to a better result, but with the back right leg of the hippo collapsing onto the belly. Our method leads to the correct registration.}
\end{figure}

This null space problem manifests itself as "shrinkage" in the registration, where regions of the source shape are pulled and scaled down to collapse degenerately to a point or sliver on the surface of the target shape, as shown didactically in \reffig{nricp_didactic}.

The popular fix to this null-space and non-smoothness problem is to use a regularizer that encourages smoothness and non-degeneracy in the registration.
Unfortunately, this regularizer ends up combating the distance objective, sacrificing the quality of the registration for an overly smooth and non-degenerate solution as shown in \reffig{reg_full_space}.

To address these issues, we depart from the closest-point projection definition of the ICP objective, and instead propose using an objective that measures the difference between the volumetric signed distance function (SDF) of the source and target shapes.
This objective sidesteps the crude closest point projection present within the ICP algorithm altogether, avoiding the shrinkage problem.
However, this objective alone is not enough to guarantee a non-degenerate solution: it opens the door for a new null-space, where the source surface may collide, twist and fold into itself however it wants, so long as the SDFs of the source and target shapes match.
\begin{figure}[t]
    \vspace*{-6mm}
  \centering
  \includegraphics[width=\linewidth]{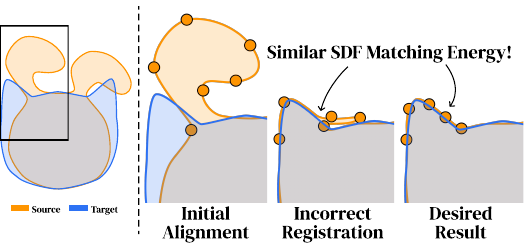}
  \caption{\label{fig:sdf_didactic}
           A didactic illustration of the null space problem in the SDF matching objective, which shows that twisting and collapsing the source surface onto itself leads to the same SDF as the true solution, so long as other parts of the source shape stretch to the target shape to compensate.}
   
\end{figure}

To avoid this new degeneracy, we propose to reparameterize our optimization problem via Skinning Eigenmodes \cite{benchekroun2023fastcomplemdynamics}, a deformation subspace for representing smooth, energy efficient, non-rigid deformations.
Introducing the subspace instead of adding a regularizer allows the optimization problem to proceed more stably without the two objectives fighting each other.
In order to resolve global transformations as well as localized deformations, we specifically propose an adaptive subspace optimization scheme, that proceeds by iteratively adding more skinning eigenmodes to the subspace and resolving the registration problem at each stage.

The result is a novel non-rigid registration algorithm that makes use of an adaptive skinning eigenmode subspace. Our method is more robust to initial conditions than standard ICP-based appraoches, capable of resolving registrations with large deformation, without the objective clash that arises between the competing objectives in existing regularized registration approaches.

We validate our approach through comprehensive experiments across diverse scenarios, including human-to-human pose transfer, large deformation registrations between different animals in different poses, and challenging cross-category registrations, and meshes with different topologies. Our evaluations demonstrate that our method outperforms traditional non-rigid ICP approaches, particularly in scenarios requiring substantial shape deformations where ICP-based methods fail. Our experiments also highlight the robustness of our method to poor mesh quality and its ability to handle cases with no clear semantic correspondence between source and target shapes.
\section{Related Work}

\subsection{Non-Rigid Registration}

The problem of registering two shapes has been studied for decades. The most classical approach to solving this problem is Iterative Closest Point (ICP), originally introduced by \cite{besl1992ogipc} for medical images. Despite decades of literature on variations to this technique, the core of the method remains largely unchanged; it iteratively computes closest-point correspondences between two shapes and estimates the rigid transformation that minimizes the distances between these correspondences.
Originally, only the Rigid Transformation that aligned one shape to another was sought, yielding Rigid-ICP(RICP). This method was widely applied in medical imaging \cite{audette2000medicalimagingsurvey}, robotics \cite{lu1994milios}, and computer graphics \cite{rusinkiewicz2001levoy} with modern variants improving computational efficiency and robustness to outliers \cite{bouaziz2013sparseclosestpoint, bouaziz2015thesis}.

While rigid registration is well-understood, non-rigid registration is far more challenging. The solution space becomes high-dimensional, and standard ICP extensions suffer from numerous local minima, null spaces and noise sensitivity \cite{sahilliouglu2020recentsurveyshapecorrespondence}. Early approaches simply adapted ICP with smoothness or rigidity priors or curvature-aware regularization \cite{amberg2007nonrigidicp, tajdari2022semicurvature,haoli2008globalcorrespondence, chui2003nonrigidicp, peng2007featurebased}. While these methods provide solutions to the non-rigid registration problem, ICP-based techniques still rely heavily on a good initial alignment and are prone to suboptimal local minima \cite{amberg2007nonrigidicp, bouaziz2016nonrigidregistrationcourse}.

A more recent shift has been toward SDF-based registration, which replaces explicit point correspondences with a continuous volumetric representation of surfaces. By aligning volumetric signed distance fields, these methods can handle noise, partial overlap, and large, complex deformations more robustly. For example, \cite{slavcheva2017killingfusion} performs non-rigid 3D reconstruction directly on SDFs, using careful regularization using Killing Fields. \cite{sellan2023RFTS} also leverages SDFs for surface reconstruction, demonstrating the advantages of a continuous, correspondence-free energy. 
While these methods alleviate problems with local-minima, they still have a large null space problem, where due to the high-dimensionality of the solution space, the method is prone to many degenerate solutions(see \reffig{sdf_didactic}).

To overcome this limitation, researchers introduced hierarchical optimization strategies. For instance, \cite{cheng2010nonrigidregistrationvectorsdf} proposes a two-stage process: first fitting a global rigid transformation, then applying locally-rigid B-spline deformations. Similarly, volumetric SDF-based methods \cite{fujiwara2011locallyrigid, zhang2015efficientvolumetric} assume that deformations are locally rigid but globally non-rigid, effectively imposing ARAP-like local-global optimization strategy \cite{sorkine2007asrigidaspossible}. These methods improve stability and reduce local minima, and remove the null space problem, but their reliance on local rigidity constraints severely limits the diversity of deformations they can capture, particularly for complex non-rigid transformations. 

Our approach builds on this line of work but goes further. Unlike prior SDF-based methods that rely on fixed local-rigidity assumptions, we build a subspace from the spectral properties of the shape that allows us to capture deformations that reflect the intrinsic geometry of the shape itself. These deformations are not constrained to be locally rigid and can often deviate significantly from rigid transformations, as demonstrated in \reffig{human_lamp} where complex bending and stretching deformations are captured. Moreover, our method naturally implements a full adaptive hierarchy, spanning from global to local deformations (see \reffig{donkey_horse}), offering finer control and expressiveness than previous two-stage or locally-rigid approaches.

\begin{figure}[t]
  \centering
  \includegraphics[width=\linewidth]{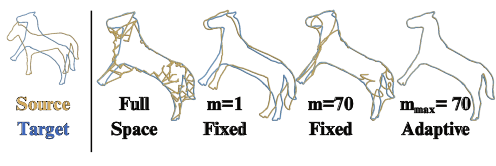}
  \caption{\label{fig:horse_2d}
          A small affine subspace can align the source and target shape globally, but cannot capture localized deformations. A higher dimensional subspace allows for localized deformations, but exhibits null-space degeneracy. Our adaptive subspace scheme gradually refines the subspace, allowing for both global and localized deformations, while avoiding the null-space degeneracy.}
\end{figure}
\begin{figure*}[t]
    \includegraphics[width=0.99\textwidth]{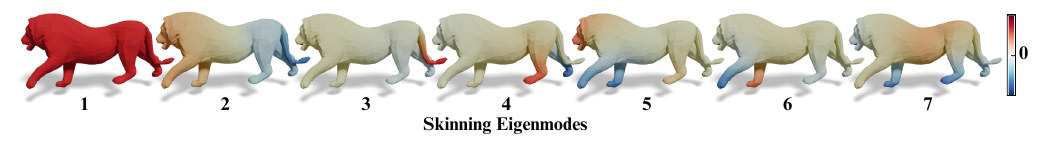}
    \centering
    \caption{We visualize the first 7 skinning eigenmodes. As they increase in number the skinning eigenmodes describe higher frequency motion, enabling for more localized deformation to occur.}
\label{fig:eigenmodes}
\end{figure*}
\subsection{Spectral Analysis for Registration}
To properly formulate our SDF-matching registration problem, we re-parameterize our degrees of freedom using the spectral properties of the shape, a long-standing theme in geometry processing.
For decades, spectral analysis has been used to approximate and smooth surface signals \cite{vallet2008Spectral}, and to build informative shape descriptors such as heat and wave-kernel signatures \cite{sun2009conciseHKS,matthieu2011WKS}.
Extending this, the functional map framework \cite{ovsjanikov2012functionalmaps} and its extensive variations, exploits the isometry-invariance of such spectral bases to construct correspondences between shapes, even shapes that are not isometrically related.
While powerful, functional maps still struggle with large non-isometric deformations, requiring user-provided landmarks or careful initialization \cite{melzi2019zoomout}.
While our method is compatible with landmarks and can also benefit from strong initialization, the explicit nature of our SDF-matching objective fundamentally differs from functional maps by directly optimizing surface alignment rather than relying on isometry-invariant correspondence functions. This allows us to find reasonable solutions even without such initialization, particularly for non-isometric deformations where functional maps break down. 

In physics-based animation, spectral methods are widely used to build low-dimensional deformation subspaces for faster simulation \cite{pentland1989goodvibrations}.
However, standard modal subspaces degrade under large rotations, motivating numerous enhancements such as rotation-strain coordinates, modal warping, and substructuring \cite{huang2011RScoordantes,choi2005modalwarping,barbic2005realtimesubspaceintegration,kim2011multidomain}.
An alternative is to construct subspaces around an animation rig, yielding linear blend skinning (LBS) subspaces that naturally accommodate large rotational motion \cite{hahn2012rigspacephysics,jacobson2012fast}.
These methods, however, depend on user-designed skinning weights or a pre-built rig.
Skinning Eigenmodes \cite{benchekroun2023fastcomplemdynamics} overcome this limitation by deriving LBS-style subspaces directly from the mesh.
They compute skinning weights via a spectral decomposition of the elastic-energy Laplacian, producing smooth, geometry- and material-aware weights without any pre-defined rig.
This automatically generated subspace is both smooth (in that deformations vary continuously across the surface) and robust to large deformations (maintaining stability even under significant rotations and stretches that would break standard modal methods).
We leverage these properties (smoothness and resilience to large motion) to re-parameterize our SDF-matching problem and efficiently solve the non-rigid registration problem. Unlike two-stage hierarchical methods that rigidly enforce local rigidity constraints, our spectral subspace naturally adapts from global to local deformations, capturing the full spectrum of possible transformations without artificial rigidity limitations.


\section{Method}
We formulate our non-rigid registration problem as an optimization problem, where we seek to find a source surface $\mat{\Gamma}$ that minimizes an objective that measures the similarity between the source and target surfaces.
\begin{align}
   \mat{\Gamma}^* =\argmin_{\mat{\Gamma}} E(\mat{\Gamma})
\end{align}
\subsection{SDF Matching Objective}
We adapt the SDF matching objective from prior work in surface reconstruction\cite{sellan2023RFTS}, seeking a surface that minimizes the difference between the SDFs of the two surfaces,
\begin{align}
    E(\mat{\Gamma}) = \int_{\Omega} ||f(\mat{p}, \mat{\Gamma}) - g(\mat{p})||^2 d\mat{p}, \label{eq:sdf_matching_continuous}
\end{align}
where $f(\mat{p}; \mat{\Gamma})$ is the signed distance function of the source surface $ \mat{\Gamma}$, evaluated at a point in space $\mat{p} \in \Rn{3}$ and $g(\mat{p})$ is the signed distance function of the target surface evaluated at that same point.

\subsection{Optimization Discretization}
We discretize our source surface via a triangle mesh with vertex position $\mat{X} \in \Rn{n \times 3}$ and triangles $\mat{T} \in \Nn{T \times 3}$ where $n$ and $T$ are the number of vertices and triangles respectively.
For convenience, we work with the vectorized form of the surface geometry, $\mat{x} \in \Rn{3n}  = \text{vec}(\mat{X})$.
We evaluate the integral in our SDF matching objective using uniform grid sampling on a set of quadrature points $\mat{Q} \in \Rn{k \times 3}$. The quadrature points are sampled uniformly on a grid covering the bounding box of both the source and target meshes. This choice provides good coverage of the domain while maintaining computational efficiency.
Our discrete non-rigid registration problem is then given by,
\begin{align}
    \mat{x}^* = \argmin_{\mat{x}} \sum_{i=1}^k ||f(\mat{Q}_i, \mat{x} ) - g(\mat{Q}_i)||^2  \label{eq:sdf_matching_discrete}
\end{align}
\subsection{Skinning Eigenmodes Subspace}
Driven by the intuition that an object is likely to deform in a way that is smooth, 
yet with localized rotations, scaling and translation, we propose to use Skinning Eigenmodes \cite{benchekroun2023fastcomplemdynamics} 
as a subspace for our non-rigid registration.
Following the Skinning Eigenmodes framework, we first compute a set of $m$ skinning weights $\mat{W} \in \Rn{n \times m}$
from the eigenanalysis of the cotangent Laplacian $\mat{L} \in \Rn{n \times n}$ of the mesh. The cotangent Laplacian and lumped mass matrices are assembled using standard formulations, given by
\begin{align}
    \mat{L} \mat{W} = \mat{M} \mat{W} \mat{\Lambda} \label{eq:skinning_eigenmodes}
\end{align}
where $\mat{M} \in \Rn{n \times n}$ is the mass matrix and $\mat{\Lambda} \in \Rn{n \times n}$ is the diagonal matrix of eigenvalues.
The resulting eigenvectors form our skinning weights $\mat{W}$.

To build our final deformation subspace, we use these skinning weights to form a Linear Blend Skinning basis \cite{jacobson2012fast},

\begin{figure}[t]
  \centering
  \includegraphics[width=\linewidth]{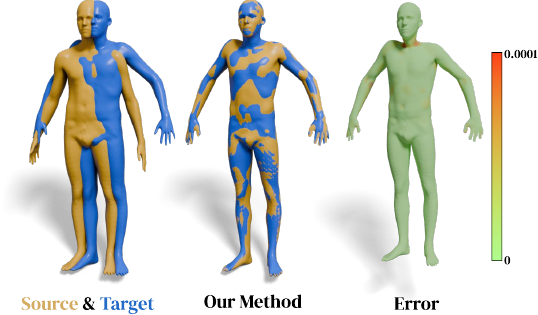}
  \caption{\label{fig:human_human}
           Our method provides registrations for humans with large differences in articulations.}
\end{figure}
\begin{figure*}[t]
  \centering
  \includegraphics[width=\linewidth]{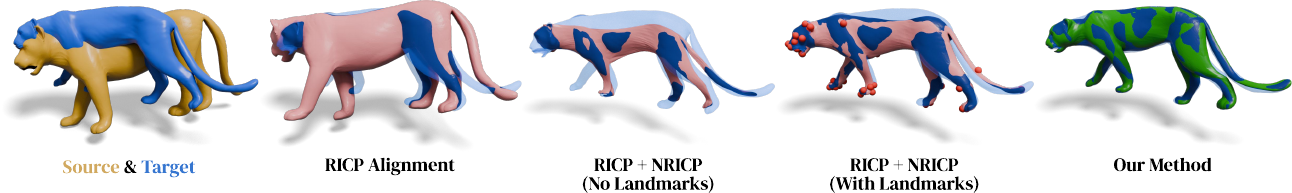}
  \caption{\label{fig:lion_cougar}
          Due to the difference in geometry between the source lion mesh and the target cougar mesh (the lion's right paw leading, in contrast with the cougar's leading right paw) RICP cannot find a good initial rigid alignment. 
          As a result, even with landmarks, NRICP results in a poor registration, collapsing the lion's back. Our method does not depend on a strong rigid alignment, and achieves a good registration without landmarks.
           }
\end{figure*}

\begin{align}
    \mat{B} = \left( \left( \mat{W} \otimes \mathbbm{1}_4^T \right) \odot \left( \mathbbm{1}_m^T \otimes \mat{\bar{X}} \right)\right) \otimes \mathbbm{I}_3 \label{eq:skinning_eigenmodes_basis}
\end{align}
where $\mat{B} \in \Rn{3n \times 12m}$ is the deformation subspace, 
$\mat{\bar{X}} \in \Rn{n \times 4}$ are the rest vertex positions in homogeneous coordinates (with the fourth coordinate being 1),$\mathbbm{1}_d \in \Rn{d \times 1}$ is a $d$-dimensional one vector and $\mathbbm{I}_3 \in \Rn{3 \times 3}$ is the identity matrix. The operators $\otimes$ denotes the Kronecker product and $\odot$ denotes the element-wise product (2D extensions of this basis, as used in \reffig{horse_2d} are also possible \cite{benchekroun2023fastcomplemdynamics}).

The full space positions of the vertices source mesh, can then be reconstructed from the low dimensional degrees of freedom $\mat{z} \in \Rn{12m}$ via,
\begin{align}
    \mat{x} = \mat{B} \mat{z} + \mat{x}_0, \label{eq:skinning_eigenmodes_reconstruciton}
\end{align}
where $\mat{x}_0 \in \Rn{3n}$ are the initial rest mesh positions.

\subsection{Adaptive Subspace Scheme}
A key advantage of using a subspace constructed from eigenanalysis is that it comes with a natural ordering for refining or coarsening the solution. The first skinning eigenmode $(m=1)$ is constant across the geometry, and as such can be used to describe global affine transformations (including translations, rotations, scale and shear). As we increase the number of skinning eigenmodes, the affine deformations are allowed to describe increasingly localized, yet smooth, deformations.

Unfortunately, the optimization problem described by \refeq{sdf_matching_discrete} is ill-posed, with many degenerate solutions that minimize the energy above.
While introducing a subspace approximation to our degrees of freedom allows us to avoid the null-space problem, low-dimensional subspaces often fail to capture localized deformations. 
In contrast, increasing the subspace dimension allows us to capture more localized deformations, but re-introduces the same null-space problem once again, allowing for degenerate solutions. This paradoxical situation is demonstrated in 2D in \reffig{sdf_didactic}.
To address this, we propose an adaptive subspace scheme that allows us to refine the subspace used throughout our optimization by adding skinning eigenmodes to the basis when the current optimization starts to stall.

We define stalling as occurring when the degrees of freedom cease to progress.
Specifically, we enrich the subspace when:
\begin{align}
   || \delta x || < \epsilon_{stall}
\end{align}
where $\delta x$ denotes the change in the degrees of freedom in one iteration. This threshold is set to start at $1e-1$ for the first subspace and is usually set to decay to $1e-3$.

For optimization, we use a gradient descent approach with a traditional backtracking line search, as detailed in \refalg{adaptive_subspace}. Convergence is determined by comparing the 2-norm of the change in deformed vertex positions between consecutive steps against a predetermined threshold. The tolerance is gradually tightened as additional subspaces are introduced.

\begin{algorithm}
\caption{Pseudocode for our adaptive morphomodes non-rigid registration algorithm. Given as input a source mesh $\mat{X} \in \Rn{n \times 3}$, a list of source mesh triangle indices $\mat{T} \in \Nn{T \times 3}$, a set of quadrature points $\mat{Q} \in \Rn{k \times 3}$, a target SDF at those quadrature points $\mat{g} \in \Rn{k \times 1}$, and a maximum number of skinning eigenmodes $m_{max}$, we output the optimized source mesh $\mat{X}^* \in \Rn{n \times 3}$.}\label{alg:adaptive_subspace}
\begin{algorithmic}
\Function{morphmodes\_registration}{$\mat{X}$, $\mat{T}$, $\mat{Q}$, $\mat{g}$, $m_{max}$}
    \State $\mat{H} \gets \text{cotangentLaplacian}(\mat{X}, \mat{T})$
    \State $\mat{M} \gets \text{massMatrix}(\mat{X}, \mat{T})$
    \State $\mat{W} \gets \text{eigs}(\mat{H}, \mat{M}, m_{max})$
    \State $\mat{B} \gets \text{buildLBSBasis}(\mat{W}, [\mat{X}, \mat{1}_n])$
    \State  $m \gets 1$
    \State  $\mat{x}_0 \gets \text{vec}(\mat{X})$
    \State  $\mat{x} \gets \mat{x}_0$
    \State  $\mat{z} \gets \text{zeros}((0, 1))$
    \While{$m < m_{max}$}
        \State $\mat{B}_m \gets \mat{B}[:,:12m]$
        \State $\mat{z} \gets \text{concatenate}([\mat{z}, \text{zeros}((12, 1))])$

        \While{$True$}
            \State $\mat{x} \gets \mat{B} \mat{z} + \mat{x}_0$
            \State $\nabla_{\mat{x}} E \gets \text{ComputeSDFGradient}(\mat{z}, \mat{x}, \mat{g}, \mat{Q})$ 
            \State $\nabla_{\mat{z}} E \gets \mat{B}^T \nabla_{\mat{x}} E$
            \State $\alpha \gets \text{lineSearch}(\mat{z}, \mat{x}, \mat{g}, \mat{Q}, \nabla_{\mat{z}} E)$
            \State $\delta z \gets - \alpha \nabla_{\mat{z}} E$
            \State $\mat{z} \gets \mat{z} + \delta z$
            \State $\delta x \gets \mat{B} \delta z$
            \If{$||\delta x|| < \epsilon_{stall}$}
                \State $\text{break}$
            \EndIf
        \EndWhile
        \State $m \gets m + 1$
        \EndWhile
        \State $\mat{X}^* \gets \text{reshape}(\mat{B} \mat{z} + \mat{x}_0, (n, 3))$
        \State \Return $\mat{X}^*$
\EndFunction
\end{algorithmic}
\end{algorithm}

\section{Experiments}
We conducted a series of experiments to demonstrate the effectiveness of our SDF-based adaptive subspace registration scheme across a variety of scenarios. Our evaluations range from classical pose transfer tasks to challenging cases involving large deformations, cross-category registrations, and meshes of poor quality. 
We also investigate the role of our adaptive subspace parameterization, showing how it enables coarse-to-fine refinement, avoids degenerate solutions that arise in full-space optimization, and eliminates the need for carefully tuned regularizers. Together, these results highlight both the robustness and flexibility of our approach. 

\emph{Pose Transfer on Human Figures}
We first evaluate our method on human-to-human registration across different poses (\reffig{human_human}). The source mesh depicts a human in a neutral standing pose, while the target mesh shows the same figure with raised arms and bent elbows. Our method produces an accurate alignment, with the deformed source mesh closely matching the target and only minimal residual error at the areas with high frequency changes.

\emph{Registration Under Large Deformations}
Our method excels in scenarios where large deformations are required to register the source mesh to the target. 

In \reffig{hippo}, we perform a registration between two hippos in different poses: the source shows a hippo walking, while the target depicts a hippo standing with its head raised upward. The deformation between these two shapes is substantial, and NRICP (non-rigid ICP) \cite{amberg2007nonrigidicp} fails to handle it reliably. When applied directly, NRICP produces severe artifacts, squishing large portions of the hippo’s body; even when combined with a RICP (rigid ICP) preprocessing step for initial alignment, it still suffers from distortions, such as flattening one of the legs. In contrast, our method produces an accurate registration that closely matches the target, without requiring rigid ICP or other alignment preprocessing. The only initialization performed is a coarse manual placement to roughly align source and target, identical to the setup used for NRICP when rigid ICP is not applied. Our method is far less sensitive to bad initial alignment than NRICP, since the SDF-based energy incorporates global shape information. By comparison, NRICP relies on closest-point correspondences that are local by construction, ignoring other unoccupied portions of the target shape, making it prone to poor local minima when the closest-point projection assigns incorrect matches.

We further demonstrate our method's ability to handle large deformations in the example of more extreme geometric differences as shown in \reffig{lion_cougar}, where the source mesh is a lion and the target mesh is a cougar, with paw placements being out of phase with each other. The initial alignment is shown on the far left, where the source and target are manually placed in roughly the same position. We then applied rigid ICP (RICP) as a preprocessing step for NRICP, but the large geometric differences between the meshes hindered RICP to produce a good initial rigid alignment. Since NRICP critically depends on a good initial alignment, it also failed to produce a reasonable registration when initialized from this RICP result. 
Even when landmarks were added, the combination of RICP and NRICP still resulted in undesirable artifacts, such as a collapse of the lion’s back. In contrast, our method achieves an accurate registration without requiring rigid ICP preprocessing or landmarks. 

\begin{figure}[t]
  \centering
  \includegraphics[width=.95\linewidth]{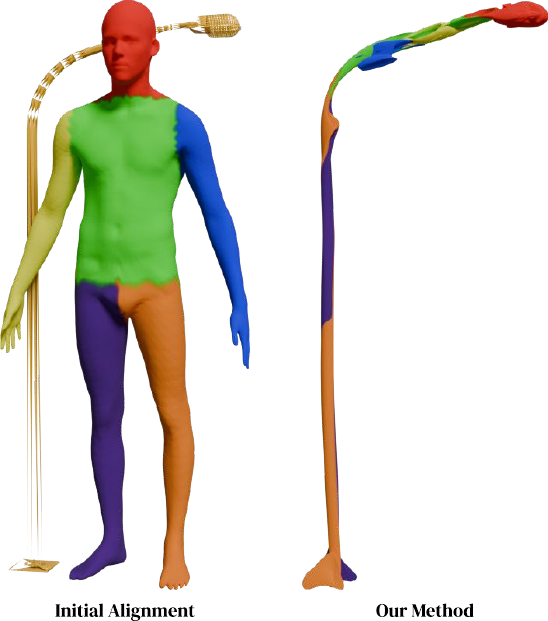}
  \caption{\label{fig:human_lamp}
           Even for extreme deformations with a non-manifold target mesh our method produces a reasonable registration, with the feet of the human mapping to the base of the lamp, and the head mapping to the top.}
\end{figure}

Additionally, to test our method on extreme non-isometric deformation, \reffig{human_lamp} shows the registration of a human mesh to a street lamp. Although such a cross-category registration has no canonical solution, our method generates a deformation that is both consistent and interpretable: the human legs naturally map to the base of the lamp, while the head aligns with the top. The resulting shape, though heavily distorted, avoids unrealistic artifacts and provides an intuitive transformation from source to target. 
This example also highlights the robustness of our approach to poor mesh quality, as the lamp mesh is non-watertight and contains numerous holes.

\begin{figure}
    \centering
    \includegraphics[width=0.95\linewidth]{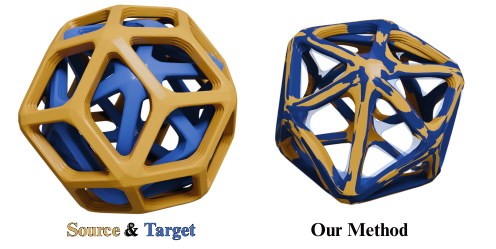}
    \caption{Our method can produce registrations between shapes of different genus.}
    \label{fig:genus_difference}
\end{figure}

Our method is also robust to differences in genus, as shown in \reffig{genus_difference}. This example registers a mesh of genus 29 to one of genus 23. The resulting solution smoothly collapses the edges of the source shape in order to match the SDF of the target shape.

\begin{figure}[htb]
  \centering
  \includegraphics[width=\linewidth]{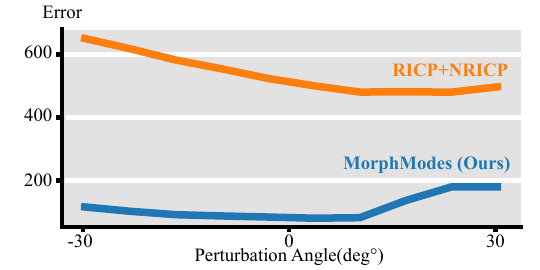}
    \caption{Our method consistently produces more accurate registrations across different perturbations on the initial orientation of the lion.}
    \label{fig:perturbation_difference}
\end{figure}

\emph{Robustness to Poor Initial Alignment}
One of the biggest advantages to using an SDF-matching energy is that the non-rigid registration becomes less sensitive to initial conditions, as it avoids the shrinkage artifacts that arise from local-minima due to the closest point projection of NRICP. 
To quantify this, \reffig{perturbation_difference} considers the registration of a lion, to that same lion in a different pose, and measures the ground truth error with respect to different initial perturbations.
Specifically, we perturb the initial orientation of the lion so that it rotates along its forward axis (with rotations varying between $-30$ and $30$ degrees), and compare registrations obtained via RICP+NRICP (NRICP initialized with RICP) and our method. We measure the error with respect to the ground truth solution, obtained from the pose transformation, and find that our method consistently achieves lower errors across the entire spectrum of initial conditions.
On average, across all initial angles tested, our method achieves an error $4.7\times$ smaller than RICP and NRICP.
\begin{figure}[t]
  \centering
  \includegraphics[width=\linewidth]{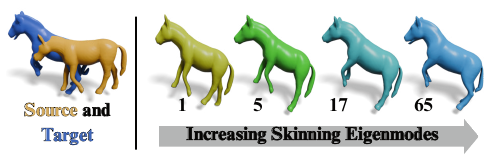}
  \caption{\label{fig:donkey_horse}
  Our adaptive optimization scheme starts with an extremely globalized alignment.
    As our adaptive registration optimization scheme progresses, it allows for increasing control over local detail, adding new degrees of freedom that describe higher frequency details.}
\end{figure}

\emph{Global-to-Local Adaptivity}
 Our adaptive subspace scheme incrementally refines the deformation by introducing higher-frequency modes as additional subspaces are added. This coarse-to-fine progression ensures that early stages already capture global structure, while later stages refine local details. \reffig{donkey_horse} illustrates this behavior on the registration of a donkey (source) to a horse (target). As more subspaces are introduced, the registration becomes increasingly accurate; nevertheless, partial registrations with limited eigenmodes are still semantically meaningful, producing mule-like shapes that reflect plausible deformations between the source and target.

\begin{figure}[t]
  \centering
  \includegraphics[width=\linewidth]{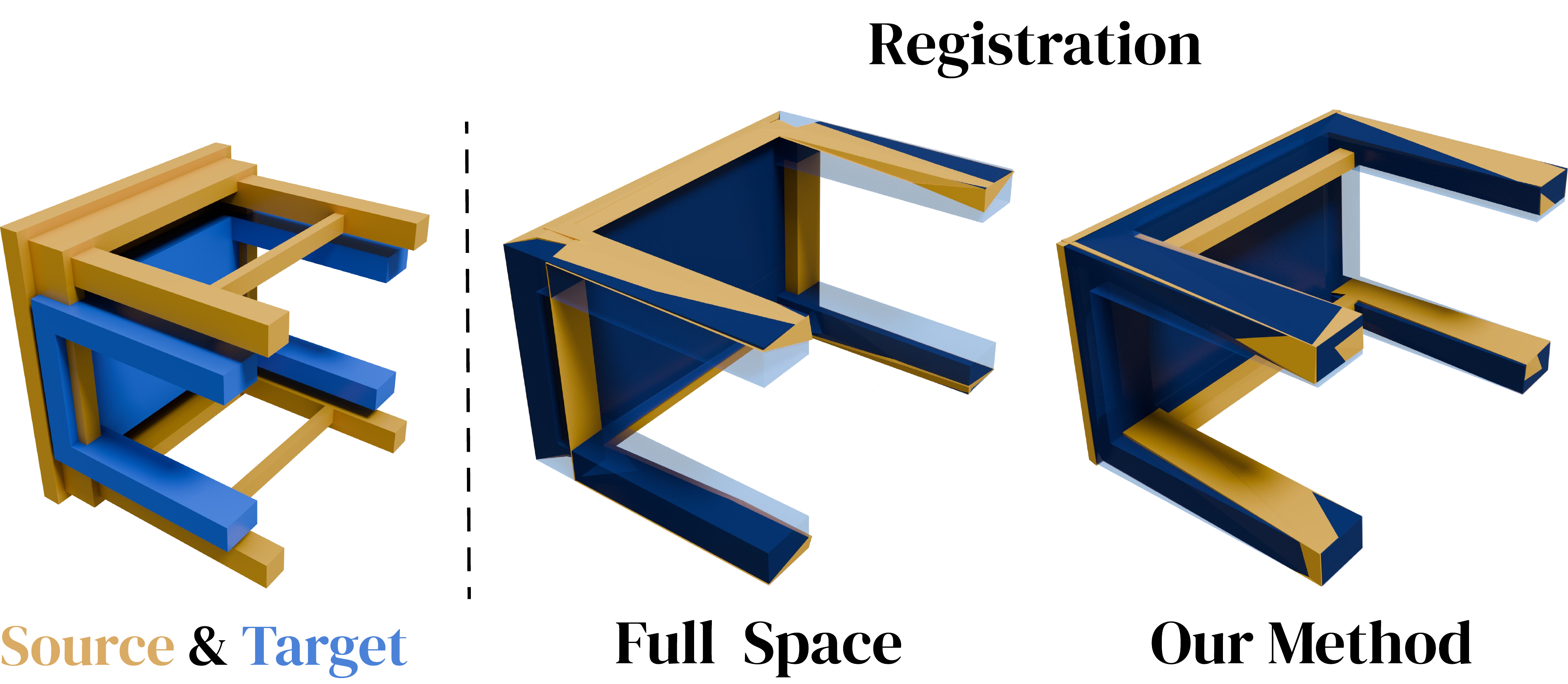}
  \caption{\label{fig:table_table}
           Even with the coarse source mesh of the table (32 vertices), degeneracies can occur when registering directly in the full space. This leads to artifacts like the collapsed leg shown above. In contrast, our adaptive subspace optimization scheme leads to a registration that more closely matches the target. }
\end{figure}

\begin{figure}[t]
  \centering
  \includegraphics[width=\linewidth]{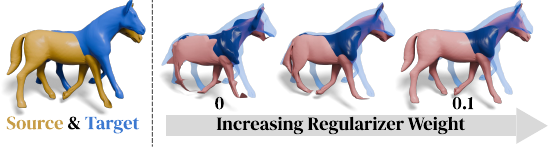}
  \caption{\label{fig:reg_full_space}
           We show that using a full-space scheme directly, leads to null-space degeneracy without a regularizer. Adding a smoothness regularizer on the displacement conflicts with the registration objective, leading to an overly stiff result.}
\end{figure}

\emph{Necessity of Adaptive Scheme}
Our registration method combines an SDF-based energy with an adaptive subspace parameterization. 

\reffig{horse_2d} walks through why an adaptive scheme is necessary for the registration problem. 
Attempting to optimize the SDF energy directly in the full vertex space of the source mesh leads to null-space degeneracies, such as branching structures and overlapping regions without volume.
Using a fixed, small subspace avoids these degenerate solutions, but comes at the cost for a partial, global alignment. 
A second alternative is to use a fixed, large subspace --- for instance, 70 modes --- but this reintroduces the null-space degeneracy artifacts.
By contrast, an adaptive scheme restricts deformations to smooth basis functions while introducing them in a coarse-to-fine manner: the optimization first captures global affine transformations before gradually adding local, higher-frequency deformations.

This global-to-local progression ensures coherent vertex motion and suppresses degenerate deformations from occurring, ensuring that by the time fine-scaled localized degrees of freedom are available, the lower-dimensional degrees of freedom that came before them had already aligned the solution to a reasonable starting point.
This result is a framework that keeps the optimization stable even under challenging conditions. 

Even when the source mesh is relatively coarse, as in the table example shown in \reffig{table_table}, these degeneracy problems can still occur. Here, direct full-space optimization causes the legs of the table to collapse, while the adaptive subspace allows for a more complete registration. 

The adaptive subspace scheme also eliminates the need for explicit regularizers in our registration. 
Regularizers are often introduced to promote smooth deformations and reduce degeneracy, but their very introduction to the problem means the optimization has to partially forego the registration objective.
This conflict, hinders convergence or preventing accurate alignment. \reffig{reg_full_space} illustrates this issue using a full-space scheme with a Dirichlet energy regularizer applied to encourage smoothness defined as
\begin{align}
  E_{dirichlet}(\mat{x}) = \frac{1}{2}(\mat{x} - \mat{x}_0)^T \mat{L} (\mat{x} - \mat{x}_0).
\end{align}

In our experiments, it proved extremely difficult to find any regularization weight that simultaneously avoided degeneracies and excessive rigidity --- values that reduced degeneracy made the deformation too stiff to align with the target, while values that allowed sufficient flexibility led to collapsed or distorted regions. This lack of a workable balance highlights the practical challenge of tuning regularizers. 

\begin{figure}[t]
  \centering
  \includegraphics[width=\linewidth]{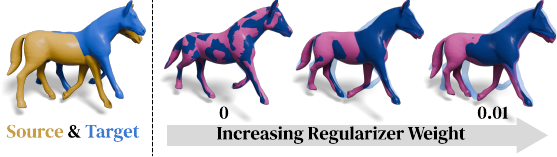}
  \caption{\label{fig:reg_subspace}
           Our adaptive subspace scheme performs well without smoothness regularization. Adding the regularizer actively conflicts with the registration objective, leading to an overly stiff result.}
\end{figure}

We further tested the same Dirichlet energy regularizer in conjunction with our adaptive subspace scheme (\reffig{reg_subspace}) and observed that our method performs \emph{better} without regularization; adding the regularizer once makes the deformation overly rigid and misaligned.
 These results demonstrate that our adaptive subspace scheme alone achieves stable and accurate registrations, without the need for regularization or parameter tuning.
\section{Limitations and Future Work}

While our method demonstrates significant improvements over existing approaches, several limitations remain. The convergence of our gradient descent optimization could be improved through more sophisticated optimization strategies, such as quasi-Newton methods or adaptive step-size schemes that better handle the changing optimization landscape as the subspace expands.
The method remains sensitive to initialization, particularly for cases where the source and target shapes have very different orientations. Poor initialization can lead to suboptimal convergence or require more iterations to achieve satisfactory results. 
The scalability to very large meshes is also a limitation, as the spectral decomposition and SDF computation scale with mesh complexity. For applications requiring real-time performance or processing of high-resolution meshes, approximate or hierarchical approaches may be necessary.

These limitations naturally open avenues for future research, including the development of more robust optimization schemes, better initialization strategies, and scalable algorithms for large-scale registration problems.

\section{Conclusion}

We have presented MorphModes, a novel approach to non-rigid registration that combines SDF-based shape matching with adaptive Skinning Eigenmode subspaces. Our method addresses fundamental limitations of existing registration techniques while providing a more robust and flexible framework for shape alignment.

Our work demonstrates the significant advantages of SDF-based energies over traditional correspondence-based methods. SDF-based energies can capture complex shape relationships that are invisible to correspondence-based approaches, enabling registrations that would be impossible with traditional ICP methods. The volumetric nature of SDFs provides a more holistic view of shape similarity that is less susceptible to local minima and more robust to poor initialization. This fundamental shift from point-wise correspondences to volumetric shape matching represents a paradigm change in how we approach non-rigid registration problems.

The adaptive subspace scheme represents our key technical contribution and reveals important insights about non-rigid registration optimization. The spectral properties of shapes provide a natural hierarchy for deformation complexity well suited for non-rigid registration—from global affine transformations captured by low-frequency modes to localized details encoded in higher-frequency modes. This hierarchy enables a principled coarse-to-fine optimization strategy that avoids degeneracy problems inherent in full-space SDF-based optimization while capturing the full spectrum of possible deformations.

Our approach demonstrates that the choice of parameterization can be more effective than regularization for controlling optimization behavior. Rather than adding competing objectives that fight against the primary registration goal, our subspace constraint naturally prevents degenerate solutions while preserving the optimization's ability to find accurate alignments. The adaptive nature of our scheme ensures that the optimization begins with simple, stable deformations and gradually introduces complexity only when needed, creating a more robust and predictable optimization process.

\bibliographystyle{eg-alpha-doi} 
\bibliography{egbibsample}     
\end{document}